\documentclass[preprint,aps,showpacs,nofootinbib]{revtex4}
\usepackage{graphicx}
\usepackage{epsfig,amssymb,amsmath,color}

\def\bea{\begin{eqnarray}}
\def\eea{\end{eqnarray}}

\begin{document}

\draft
\tighten
\preprint{
CTPU-14-01, DESY-14-051
}
\title{\Large \bf
String theoretic QCD axions \\
in the light of PLANCK and BICEP2
}
\author{
Kiwoon Choi$^{\,a,b}$\footnote{e-mail: kchoi@kaist.ac.kr},
Kwang Sik Jeong$^{\,c}$\footnote{e-mail: kwangsik.jeong@desy.de},
Min-Seok Seo$^{a}$\footnote{e-mail: minseokseo@ibs.re.kr}
 }
\affiliation{$^a$ $\mbox{\it Center for Theoretical Physics of the Universe,
     IBS, Daejeon 305-811, Korea}$
     \\
     $^b$ $\mbox{\it Department of Physics, KAIST, Daejeon, 305-701, Korea}$
     \\
     $^c$ 
     DESY, Notkestrasse 85, 22607 Hamburg, Germany
    }


\begin{abstract}

The QCD axion solving the strong CP problem may originate from antisymmetric tensor gauge
fields in compactified string theory, with a decay constant around the GUT scale.
Such possibility appears to be ruled out now by the detection of tensor modes by BICEP2 and
the PLANCK constraints on isocurvature density perturbations.
A more interesting and still viable possibility is that the string theoretic QCD axion is
charged under an anomalous U$(1)_A$ gauge symmetry.
In such case, the axion decay constant can be much lower than the GUT scale if moduli are
stabilized near the point of vanishing Fayet-Illiopoulos term, and U$(1)_A$-charged matter
fields get a vacuum value $v\sim (m_{\rm SUSY}M_{Pl}^n)^{1/(n+1)}$ $(n\geq 0)$ induced by
a tachyonic SUSY breaking mass $m_{\rm SUSY}$.
We examine the symmetry breaking pattern of such models during the inflationary epoch with
$H_I\simeq 10^{14}$~GeV, and identify the range of the QCD axion decay constant, as well as
the corresponding relic axion abundance, consistent with known cosmological constraints.
In addition to the case that the PQ symmetry is restored during inflation, i.e. $v(t_I) =0$,
there are other viable scenarios, including that the PQ symmetry is broken during inflation
with $v(t_I) \sim (4\pi H_IM_{Pl}^n)^{1/(n+1)}\sim 10^{16}$--$10^{17}$~GeV
due to the Hubble-induced $D$-term $D_A\sim 8\pi^2 H_I^2$,
while $v(t_0)\sim (m_{\rm SUSY}M_{Pl}^n)^{1/(n+1)}\sim 10^{9}$--$5\times 10^{13}$~GeV
in the present universe, where $v(t_0)$ above $10^{12}$~GeV requires a fine-tuning of
the axion misalignment angle.
We also discuss the implications of our results for the size of SUSY breaking soft masses.

\end{abstract}

\maketitle

\section{Introduction and summary}\label{sec:1}

The strong CP problem~\cite{axion-review} of the Standard Model of particle physics is about
the question why the strong CP violating parameter
$\bar\theta =\theta_{\rm QCD}+\arg(y_uy_d)$ is smaller than $10^{-10}$,
while the weak CP violating Kobayashi-Maskawa phase originating from the same quark Yukawa
couplings $y_{u,d}$ is of order unity. Presently the most compelling solution to this problem
is to introduce a non-linearly realized anomalous global U$(1)$ symmetry, the Peccei-Quinn
(PQ) symmetry~\cite{Peccei:1977hh}, which predicts a pseudo-Goldstone boson, the QCD axion,
whose vacuum expectation value (VEV) can be identified as
$\bar\theta$~\cite{old-axion-papers,KSVZ,DFSZ}.
Yet, there still remain some questions.
One question is, what is the origin of the PQ symmetry?
The PQ symmetry is required to be explicitly broken by the QCD anomaly, while being protected
well from other forms of explicit breaking.
In view of that global symmetry is not respected in general by UV physics at scales where
quantum gravity becomes important~\cite{global-symmetry-gravity}, the existence of such global
symmetry at low energy scales may require a specific form of UV completion of
the model~\cite{UV-PQ-symmetry}.
Another question is about the mechanism to determine the axion decay constant $f_a$,
which determines most of the phenomenological consequences of the QCD axion, including
the cosmological ones.

It has been known for many years that string theory provides an attractive theoretical
framework to address these questions~\cite{Witten:1984dg}.
String theory includes a variety of higher-dimensional antisymmetric tensor gauge fields,
whose zero modes behave like axions in the 4-dimensional effective theory.
The shift symmetries associated with these axion-like fields are valid in perturbation
theory~\cite{axion-string-theory,Ibanze-Uranga}.
It is then conceivable that a certain combination of the shift symmetries is broken dominantly
by the QCD anomaly, and therefore can be identified as the PQ symmetry solving the strong
CP problem.
As for the decay constant, if the compactification scale is comparable to the Planck scale,
the decay constants of such stringy axions are estimated
to be~\cite{Choi:1985bz,Svrcek:2006yi,axion-models-in-string},
\bea
\label{axion_scale1}
f_a \,\sim\, {g^2M_{Pl}}/{8\pi^2},
\eea
where the factor $8\pi^2$ comes from the convention for the axion decay constant,
and $M_{Pl} \simeq 2.4\times 10^{18}$~GeV is the reduced Planck scale.
Although it is subject to severe cosmological
constraints~\cite{axion-cosmology1,works-on-axion-iso,axion-cosmology2},
such QCD axion arising from antisymmetric tensor gauge fields
in compactified string theory has been considered to be a viable possibility for many years.

An interesting generalization of this scheme, involving an anomalous U$(1)_A$ gauge symmetry
with a nonzero U$(1)_A$-SU$(3)_c$-SU$(3)_c$ anomaly cancelled by the 4-dimensional
Green-Schwarz (GS) mechanism~\cite{Green:1984sg}, has been discussed before for the purpose
of having an intermediate scale QCD axion even when the compactification scale is comparable
to the Planck scale~\cite{Barr:1985hk,Svrcek:2006yi,Choi:2011xt}.
It is based on the compactification models in which moduli are stabilized at the point of
vanishing U$(1)_A$ Fayet-Illiopoulos (FI) term $\xi_{\rm FI}=0$ in the supersymmetric limit,
when all U$(1)_A$-charged matter fields $\phi$ are set to zero.
Such supersymmetric solutions are known to exist in many of the Type II string theory with
$D$-branes~\cite{vanishing-FI,Ibanze-Uranga}, as well as in the heterotic string theory
with U$(1)$ gauge bundles~\cite{Blumenhagen:2005ga,Anderson:2009nt}.
In the limit of $\xi_{\rm FI}=\phi=0$, the U$(1)_A$ gauge boson obtains a superheavy mass
$M_A \sim M_{Pl}/8\pi^2$ by absorbing the stringy axion $\theta_{\rm st}$
implementing the GS anomaly cancellation mechanism,
while leaving an unbroken perturbative global U$(1)$ symmetry, which corresponds the global
part of U$(1)_A$ without the transformation of $\theta_{\rm st}$.
By construction, this perturbative global U$(1)$ symmetry has nonzero
U$(1)$-SU$(3)_c$-SU$(3)_c$
anomaly, and therefore can be identified as the PQ symmetry solving the strong CP problem.

To satisfy the astrophysical constraints on the QCD axion, this PQ symmetry should be
spontaneously broken at a scale higher than $10^{9}$~GeV~\cite{axion-review}.
For this, some U$(1)_A$-charged matter  field $\phi$ should have a tachyonic supersymmetry
(SUSY) breaking scalar mass $m_{\rm SUSY}$, destabilizing the supersymmetric solution
$\xi_{\rm FI}=\phi=0$.
The matter  scalar field  $\phi$ then takes a vacuum value $\langle \phi\rangle > 10^9$~GeV
by an interplay between the tachyonic SUSY breaking mass term and a supersymmetric higher
order term  which schematically takes the form
$|\phi|^{2n+4}/M_{Pl}^{2n}$ with $n\geq 0$ if the cutoff-scale of the model is assumed to
be comparable to the Planck scale~\cite{fa-susy}.
This scheme to determine $\langle \phi\rangle$ leads to an appealing connection between
the axion scale and the SUSY breaking scale as
\bea\label{axion_scale2}
f_a \,\simeq \,\langle \phi\rangle \,\sim\, (m_{\rm SUSY} M_{Pl}^n)^{1/(n+1)}
\quad (n\geq 0),
\eea
which makes it possible that a wide range of the QCD axion decay constant much lower than
the Planck scale is obtained within the framework of string theory.

The recent detection of tensor modes in the cosmic microwave background (CMB) by
BICEP2~\cite{Ade:2014xna} has important implications for axion cosmology~\cite{axion-after-bicep2},
particularly for the string theoretic QCD axion.
First of all, the BICEP2 results imply that the inflation energy scale is about $10^{16}$~GeV.
This suggests that the string compactification scale is higher than $10^{16}$~GeV, and therefore
the estimate (\ref{axion_scale1}) of the decay constants of stringy axion-like fields is
at least qualitatively correct.
For the expansion rate $H_I \sim 10^{14}$~GeV, if the PQ symmetry were spontaneously broken during
inflation, the corresponding QCD axion is severely constrained by the PLANCK constraints
on isocurvature density perturbations and non-Gaussianity~\cite{Ade:2013uln}.\footnote{
It is in principle possible that the axion under the consideration obtains a heavy mass
$m_a(t_I) \gtrsim H_I$ during inflation, so is free from the isocurvature and non-Gaussianity
constraints~\cite{Jeong:2013xta,Higaki:2014ooa}.
However, it is not likely to be realized in our theoretical framework,
as $m_a$ is protected by both the shift symmetry broken only by non-perturbative effects and
the softly broken SUSY during inflation with $H_I \ll M_{Pl}$.
}
As we will see, this rules out the simple possibility that the QCD axion corresponds
to a combination of the zero modes of antisymmetric tensor fields in compactified string theory,
having a decay constant $f_a\sim g^2M_{Pl}/8\pi^2$.
On the other hand, in the presence of an anomalous U$(1)_A$ gauge symmetry with vanishing FI term,
under which the QCD axion is charged, the model can have rich symmetry breaking patterns during
inflation, while giving a present axion decay constant much lower than $g^2M_{Pl}/8\pi^2$.
This may make it possible that the model allows a variety of different cosmologically viable
scenarios.

In this paper, we examine the symmetry breaking pattern of the string theoretic QCD axion
models involving an anomalous U$(1)_A$ gauge symmetry during the inflationary epoch with
$H_I\simeq 10^{14}$~GeV.
We identify the allowed range of the axion decay constant in such models, as well as
the corresponding relic axion abundance, being consistent with  known cosmological
constraints, within a general framework in which the axion scale
during inflation can be different from the axion scale in the present universe.
We note first that if the PQ symmetry were broken during inflation, the cosmological
constraints can be satisfied {\it only} when the axion scale during inflation is much
higher than the present axion scale.
The most natural setup to realize this possibility is to generate the axion scale through
SUSY breaking effects.
We show that indeed the string theoretic QCD axion models with anomalous U$(1)_A$ gauge
symmetry provides such setup.
If the modulus-axion superfield implementing the GS mechanism is {\it not} sequestered
from the SUSY breaking by the inflaton sector, which would be the case in generic situations,
U$(1)_A$-charged matter fields develop a large expectation value during
inflation,
\bea
\langle \phi(t_I)\rangle \sim (\sqrt{8\pi^2} H_I M_{Pl}^n)^{1/(n+1)},
\eea
due to the tachyonic SUSY breaking scalar mass induced dominantly by the $U(1)_A$ $D$-term:
\bea
m^2_\phi(t_I) \simeq q_\phi g^2_A D_A(t_I) \sim -8\pi^2 H_I^2,
\nonumber
\eea
while
\bea
\langle \phi(t_0)\rangle \sim (m_{\rm SUSY}M_{Pl}^n)^{1/(n+1)},
\nonumber
\eea
for the SUSY breaking scalar mass $m_{\rm SUSY}$ in the present universe.
Then the QCD axion during inflation has a much higher decay constant than the present value,
and even is a different degree of freedom.
As we will see, this makes it possible that a certain parameter space of the model is
consistent with the constraints on isocurvature perturbations and non-Gaussianity,
as summarized in Fig.~\ref{fig:axion} in section \ref{sec:3}.
The allowed range of the present axion decay constant for reasonable choice of model
parameters is given by
\bea
10^{9} \,\, {\rm GeV} \,\lesssim\, f_a(t_0) \,\lesssim\, 5\times 10^{13} \,\, {\rm GeV},
\eea
where $f(t_0)\gtrsim 10^{12}$~GeV requires a fine-tuning of the axion misalignment angle
as $\theta_0\lesssim {\cal O}(10^{-1})$.
If we assume $\theta_0={\cal O}(1)$, the allowed range is reduced to
$f_a(t_0)\simeq 10^9$--$10^{11}$~GeV, with the axion dark matter making up
roughly $0.1$--$10$~\% of the total dark matter energy density.

On the other hand, if the modulus-axion superfield for the GS mechanism is sequestered
from the SUSY breaking by the inflaton sector, so that the soft scalar masses during inflation
are not dominated by the U$(1)_A$ $D$-term contribution, it is possible that
\bea
\langle \phi(t_I)\rangle =0,
\eea
so the PQ symmetry is restored during inflation,
while again
\bea
\langle \phi(t_0)\rangle \sim (m_{\rm SUSY}M_{Pl}^n)^{1/(n+1)},
\nonumber
\eea
in the present universe.
In this case, the model is free from the isocurvature and non-Gaussianity constraints,
however required to have the axion domain-wall number $N_{\rm DW}=1$, which is a non-trivial
constraint on the model building.
Furthermore, if one adopts the recent simulation for the axion production
by axionic strings and domain walls~\cite{Hiramatsu:2012gg}, only the following narrow window
of the axion decay constant
\bea
10^9~{\rm GeV}
\,\lesssim\, f_a(t_0)\, \lesssim\, (\mbox{a few})\times 10^{10}~{\rm GeV}
\eea
is allowed by the astrophysical and cosmological constraints, where
the relic axions can account for the total dark matter energy density
when $f_a(t_0)$ saturates the upper bound.

Our results have an intriguing implication for the size of SUSY breaking soft masses
in the present universe.
Regardless of whether the PQ symmetry is broken or not during inflation, the cosmologically allowed
parameter region for a natural axion misalignment angle $\theta_0={\cal O}(1)$
points to two possibilities:\footnote{
The possibility of the axion scale SUSY was noticed also in Ref.~\cite{Hall:2014vga}
recently.}
\bea
&i)& \mbox{Axion scale SUSY:}\quad  m_{\rm SUSY}\,\sim\,
f_a(t_0) \,\sim\, 10^9-10^{11}~{\rm GeV},
\nonumber \\
&ii)& \mbox{Low scale SUSY:}\quad  m_{\rm SUSY}\,\sim\,
f_a^2(t_0)/M_{Pl} \,\sim \,10^3-10^4~{\rm GeV}.
\eea
The results for the case of broken PQ symmetry during inflation suggest also
that the axion isocurvature density perturbations have an amplitude close
to the present observational bound.

The organization of this paper is as follows.
In section \ref{sec:2}, we review the relevant features of the string theoretic QCD axion.
In section \ref{sec:3}, we examine the cosmological constraints on the QCD axion,
while taking into account that the axion decay constant during inflation can be much higher
than the present value.
Although we consider here a specific type of string motivated models, it should be noted that
our results apply to generic supersymmetric axion models in which the PQ breaking scale is
generated by SUSY breaking effects.
In section \ref{sec:fa}, we present a simple 4-dimensional supergravity (SUGRA) model
involving both the inflaton sector and the U$(1)_A$ sector, and examine possible symmetry
breaking patterns during inflation.

\section{String theoretic QCD axion}\label{sec:2}

String theory contains a variety of higher-dimensional antisymmetric $p$-form gauge fields
$C_p$, together with the associated gauge symmetry, under which
\bea
\label{gauge1}
C_p \rightarrow C_p + d\Lambda_{p-1},
\eea
where $\Lambda_{p-1}$ is
a $(p-1)$-form parameterizing the gauge transformation.
For compactifications involving a $p$-cycle $\alpha_p$ in the internal space,
the resulting 4-dimensional effective theory contains an  axion-like field $\theta_{\rm st}$:
\bea
C_p(x,y) = \theta_{\rm st}(x) \omega_p(y),
\eea
where $x$ and $y$ denote the coordinates of the 4-dimensional flat Minkowski spacetime
and the internal space, respectively,
and $\omega_p$ is a harmonic $p$-form with $\int_{\alpha_p} \omega_p =1$.
Since $\omega_p(y) = d\Omega_{p-1}(y)$ locally,
the shift symmetry
\bea
\label{shift}
U(1)_{\rm shift}: \,\, \theta_{\rm st}(x) \rightarrow \theta_{\rm st}(x)+{\rm constant}
\eea
is locally equivalent to the gauge symmetry (\ref{gauge1}), but not globally
due to the obstruction from $\int_{\alpha_p} \omega_p \neq 0$.
This implies that the shift symmetry (\ref{shift}) is valid in perturbation theory,
but can be  broken by non-perturbative effects associated with
$\int_{\alpha_p}\omega_p \neq 1$.
Such effects include for instance the stringy-instantons wrapping $\alpha_p$,
as well as the axion couplings to the low energy gauge field instantons, which are induced as
\bea
\int C_p\wedge F\wedge F
\quad \rightarrow \quad \int_{M_4} \theta_{\rm st}F\wedge F \int_{\alpha_p} \omega_p.
\eea
It is then a conceivable possibility that stringy instanton effects are negligible
for the shift symmetry (\ref{shift}), at least in the limit that the $p$-cycle $\alpha_p$
is large enough.
This would allow that the shift symmetry (\ref{shift}) is explicitly broken
{\it dominantly} by the QCD anomaly, and so the stringy axion $\theta_{\rm st}$ can
be identified as the QCD axion solving the strong CP problem.

A characteristic feature of such string theoretic axion is that
its decay constant is of the order of $M_{Pl}/8\pi^2$ if the compactification scale
is comparable to the Planck scale, where $8\pi^2$ is a conventional factor for the axion
decay constant.
To see this, one can consider the 4-dimensional effective SUGRA of the modulus-axion
superfield
\bea
T = \frac{1}{2}\tau + i\theta_{\rm st},
\eea
where $\tau$ is the modulus partner of $\theta_{\rm st}$, describing the volume
of the $p$-cycle $\alpha_p$.
For the modulus K\"ahler potential $K_0$ and the holomorphic gauge kinetic function
$\tilde f_\alpha$ for the QCD, which generically take the form,
\bea
K=K_0(T+T^*), \quad \tilde f_\alpha = T + \cdots,
\eea
the effective lagrangian of $\theta_{\rm st}$ reads
\bea
{\cal L}_{\rm eff} &=&
M_{Pl}^2 \frac{\partial^2 K_0}{\partial \tau^2}\partial_\mu \theta_{\rm st}
\partial^\mu \theta_{\rm st}
+ \frac{1}{4} \theta_{\rm st}G^{\alpha\mu\nu}\tilde G^\alpha_{\mu\nu}+ \cdots
\nonumber \\
&=& \frac{1}{2}\partial_\mu a_{\rm st} \partial^\mu a_{\rm st}
+ \frac{1}{32\pi^2}\frac{a_{\rm st}}{f_a}G^{a\mu\nu}\tilde G^a_{\mu\nu}+ \cdots,
\eea
where $a_{\rm st}$ is the canonically normalized string theoretic QCD axion,
$G^\alpha_{\mu\nu}$ is the gluon field strength, and the axion decay constant is given by
\bea
f_a = \frac{1}{8\pi^2}\left(2\frac{\partial^2 K_0}{\partial \tau^2}\right)^{1/2}M_{Pl}.
\eea
The BICEP2 results imply that the inflation energy scale is about $10^{16}$~GeV, and therefore
the compactification scale is higher than $10^{16}$~GeV.
Such a high compactification scale implies that the modulus
K\"ahler metric ${\partial^2 K_0}/{\partial \tau^2}$ is {\it not} significantly  smaller
than the unity.
More specifically, from the QCD gauge kinetic function which depends on $T$, and thereby suggests
$\tau\sim 1/g_{\rm GUT}^2$, the modulus K\"ahler metric can be estimated as
\bea
\left(\frac{\partial^2 K_0}{\partial \tau^2}\right)^{1/2} \,=\,{\cal O}(g^2_{\rm GUT}).
\eea
This gives
\bea
f_a\, =\, {\cal O}\left(g_{\rm GUT}^2M_{Pl}/8\pi^2\right)\label{decay1},
\eea
which turns out to be a correct estimate for the most of compactification models\footnote{
One may be able to obtain a much lower axion scale, while keeping the cutoff-scale
for the inflaton sector higher than $10^{16}$~GeV, if the axion sector and the inflaton
sector are separated from each other in a warped internal space~\cite{Choi:2003wr}.
Here we do not pursue this kind of more involved possibility.
}
with a compactification scale higher than $10^{16}$~GeV.
It has been known for many years that this type of string theoretic QCD axion is subject
to severe cosmological constraints.
As we will see in section \ref{sec:3}, it appears to be ruled out now by the detection
of tensor modes by BICEP2 and the PLANCK constraints on isocurvature density perturbations.

In fact, the QCD axion can have a decay constant far below $M_{Pl}/16\pi^2$ even when
the compactification scale is comparable to the Planck scale.
An attractive scheme to realize such possibility is that the stringy axion
$\theta_{\rm st}$ is charged under an anomalous U$(1)_A$ gauge symmetry,
and its modulus partner $\tau$ is
stabilized at a value near the point of vanishing FI term.\footnote{
See Ref.~\cite{Cicoli:2013cha} for string axions with vanishing FI term in the large volume scenario.
}
Indeed, such scheme can be realized in many string compactification models, including
the Type II string models with $D$-branes and the heterotic string models with U$(1)$
Yang-Mills bundles on Calabi-Yau manifold.
Four-dimensional symmetries of this type of models include a shift symmetry
\bea
U(1)_{\rm shift}:\quad
T\rightarrow T+i c \quad (c=\mbox{real constant}),
\eea
which is broken dominantly by the QCD anomaly, as well as an anomalous $U(1)_A$ gauge symmetry:
\bea
U(1)_A: \quad V_A \rightarrow V_A+\Lambda+\Lambda^*, \quad
T\rightarrow T+\delta_{\rm GS}\Lambda, \quad
\phi_i \rightarrow e^{q_i\Lambda}\phi_i,
\eea
where $V_A$ is the vector superfield for the U$(1)_A$ gauge multiplet,
$\phi_i$ are generic $U(1)_A$-charged chiral matter superfields,
$\Lambda$ is a chiral superfield parameterizing U$(1)_A$ transformation on the superspace,
and
\bea
\label{GS}
\delta_{\rm GS}=\frac{1}{8\pi^2}\sum_i q_i {\rm Tr}(T_a^2(\phi_i))
\eea
represents the coefficient of the mixed U$(1)_A$-SU$(3)_c$-SU$(3)_c$ anomaly which is
cancelled by the GS mechanism.

Generically the K\"ahler potential and the QCD gauge kinetic function take the form,
\bea
K &=& K_0(T+T^*-\delta_{\rm GS}V_A) + Z_i(T+T^*-\delta_{\rm GS}V_A)
\phi_i^* e^{-q_i V_A} \phi_i
+ \cdots,
\nonumber \\
\tilde f_\alpha &=& T + \cdots.
\eea
In the following, for simplicity, we will consider only a single U$(1)_A$-charged matter
field $\phi$ under the assumption that its K\"ahler metric is a moduli-independent constant.
Then the relevant part of the effective lagrangian is given by
\bea
{\cal L}_{\rm eff} &=&-\frac{1}{4g^2_A}F^{\mu\nu}F_{\mu\nu}
+ M_{Pl}^2\frac{\partial^2 K_0}{\partial \tau^2} \left(\partial_\mu \theta_{\rm st}
- \delta_{\rm GS}A_\mu\right)^2
+ D_\mu \phi^* D^\mu \phi
\nonumber \\
&&
+\,
\frac{1}{2} g^2_A\left(\delta_{\rm GS}\frac{\partial K_0}{\partial \tau}
- |\phi|^2\right)^2+\frac{1}{4} \left(\theta_{\rm st}
- \delta_{\rm GS}\arg(\phi)\right)G^{a\mu\nu}\tilde G^a_{\mu\nu}+ \cdots,
\eea
where we have set $Z_\phi=q_\phi=1$, and included the counter term for the mixed
U$(1)_A$-SU$(3)_c$-SU$(3)_c$ anomaly, whose U$(1)_A$ variation is cancelled by
the gauge variation of $\theta_{\rm st}$.
The above effective lagrangian can be rewritten as
\bea
{\cal L}_{\rm eff}&=&
-\frac{1}{4g^2_A}F^{\mu\nu}F_{\mu\nu}+\frac{1}{2}
\left((8\pi^2\delta_{\rm GS} f_{\rm st})^2+v^2 \right)
\Big( \frac{\partial_\mu \chi}{\sqrt{ (8\pi^2\delta_{\rm GS} f_{\rm st})^2 +v^2}}-A_\mu\Big)^2
\nonumber \\
&& +\,
\frac{1}{2}(\partial_\mu a)^2
+ \frac{1}{32\pi^2} \frac{a}{f_a}\,G\tilde G
+\frac{1}{2} g^2_A (\xi_{\rm FI}-v^2/2)^2+\cdots,
\eea
for $\theta_\phi =\arg(\phi)$, and $v=\sqrt2\langle \phi \rangle$.
Here $\chi$ and $a$ are given by
\bea
\chi &=&
\frac{1}{\sqrt{(8\pi^2\delta_{\rm GS} f_{\rm st})^2+v^2}}
\left(
(8\pi^2\delta_{\rm GS} f_{\rm st})^2 \frac{\theta_{\rm st} }{\delta_{\rm GS}}
+ v^2 \theta_\phi \right),
\\
\label{canonical-axion}
a &=&
\frac{(8\pi^2\delta_{\rm GS} f_{\rm st}) v}{\sqrt{(8\pi^2\delta_{\rm GS} f_{\rm st})^2+v^2}}
\left(\frac{\theta_{\rm st} }{\delta_{\rm GS}}- \theta_\phi\right),
\eea
with $f_{\rm st}$ and $f_a$ defined by
\bea
\label{fa-axion}
f_{\rm st} &\equiv& \frac{1}{8\pi^2}\sqrt{2\frac{\partial^2 K_0}{\partial \tau^2}}M_{Pl},
\nonumber \\
f_a &\equiv&
\frac{f_{\rm st} v}{\sqrt{(8\pi^2\delta_{\rm GS} f_{\rm st})^2+v^2}}.
\eea
Note that the U$(1)_A$ $D$-term includes the moduli-dependent FI term,
\bea
\xi_{\rm FI} = \delta_{\rm GS}\frac{\partial K_0}{\partial \tau} M_{Pl}^2.
\eea
Obviously $\chi$ corresponds to the longitudinal component of
the massive U$(1)_A$ gauge boson with a mass
\bea
M_A = g_A \sqrt{(8\pi^2\delta_{\rm GS} f_{\rm st})^2 + v^2},
\eea
while $a$ is the physical QCD axion and $f_a$ is its decay constant.
When the compactification scale is higher than $10^{16}$~GeV, the modulus K\"ahler metric
typically has a vacuum value as
$\langle \partial^2 K_0/\partial\tau^2 \rangle \sim 1/\langle \tau \rangle^2$,
and the gauge coupling constant is given by
$1/g^2_{\rm GUT} = \langle \tau \rangle/2 + \cdots$.
Thus $f_{\rm st}$ is around $g^2_{\rm GUT} M_{Pl}/8\pi^2$, or it may be possible to
increase it by one order of magnitude \cite{Svrcek:2006yi}, implying
\bea
f_{\rm st} = {\cal O}(10^{-1} - 10^{-2})\times M_{Pl},
\eea
independently of the details of moduli stabilization.

On the other hand, the matter vacuum expectation value $v=\sqrt2\langle \phi\rangle$
severely depends on the mechanism of moduli stabilization, particularly on
the vacuum value of the FI term.
In 4-dimensional $N=1$ SUGRA with $m_{3/2}\ll M_A$ for $m_{3/2}$ being the gravitino mass,
we have the following bound on the $D$-term:
\bea
|D_A| = |\xi_{\rm FI}-v^2/2|
\lesssim {\cal O}\left(\frac{m_{3/2}^2M_{Pl}^2}{M_A^2}\right),
\eea
which can be derived from the stationary condition for the scalar potential~\cite{D-term-size}.
Then there are essentially two distinctive possibilities.
One is that the modulus $\tau$ is stabilized at a value with
\bea
\frac{\partial K_0}{\partial \tau} ={\cal O}(1),
\eea
which is the case, for instance, when $\theta_{\rm st}$ is the model-independent
axion and $\tau$ is the dilaton in the heterotic string theory.
In this case, we have
\bea
\xi_{\rm FI} \,\simeq \, v^2 \,=\, {\cal O}(\delta_{\rm GS}M_{Pl}^2)\, > \,
f_{\rm st}^2 \,=\, {\cal O}(M_{Pl}^2/(8\pi^2)^2).
\eea
Then the physical QCD axion is mostly $\theta_{\rm st}$ which originates from
antisymmetric tensor gauge fields, and its decay constant reads
\bea
f_a \,=\,
\frac{f_{\rm st} v}{\sqrt{(8\pi^2\delta_{\rm GS} f_{\rm st})^2+v^2}}
\,\simeq\, f_{\rm st}.
\eea
Axion cosmology in this case is essentially the same as in the case without
anomalous U$(1)_A$ symmetry, and therefore the model is in conflict with
the inflation scale $H_I\simeq 10^{14}$~GeV.

Another, more interesting, possibility is that the modulus $\tau$ is stabilized
at a value near the point of vanishing FI-term.
Most of the known models with anomalous U$(1)_A$ symmetry, realized either in
the Type II string theory with $D$-branes or in the heterotic string theory with U$(1)$
gauge bundles, admit a supersymmetric solution with
\bea
\xi_{\rm FI} = \phi =0.
\eea
To be phenomenologically viable, this solution should be destabilized by a
{\it tachyonic} SUSY breaking mass of $\phi$ to develop  $v>10^9$~GeV.
Schematically the scalar potential of $\phi$ takes the form
\bea
V(\phi)= -m_{\rm SUSY}^2 |\phi|^2
+ \left|\frac{\partial W}{\partial \phi}\right|^2
=
-m_{\rm SUSY}^2 |\phi|^2 + \frac{1}{M_{Pl}^{2n}}|\phi|^{4+2n} \quad (n\geq 0),
\eea
yielding
\bea
v \,\sim \,\left(m_{\rm SUSY} M_{Pl}^n\right)^{1/(n+1)} \,\ll\, f_{\rm st},
\eea
where the SUSY breaking mass $m_{\rm SUSY}$ is assumed to be small
enough compared to $M_{Pl}$.
In this case, the physical QCD axion is mostly $\theta_\phi=\arg(\phi)$,
and the axion decay constant is determined by $v$,
\bea
f_a \,=\,
\frac{f_{\rm st} v}{\sqrt{(8\pi^2\delta_{\rm GS} f_{\rm st})^2+v^2}}
\,\simeq\,
\frac{v}{8\pi^2\delta_{\rm GS}},
\eea
where $8\pi^2\delta_{\rm GS}=\sum_i q_i{\rm Tr}(T^2_a(\phi_i))$,
and we have set $q_\phi=1$.

So far, we have discussed the axion decay constant in the present universe in models
with anomalous U$(1)_A$ gauge symmetry.
An interesting feature of the axion models discussed above, providing an intriguing
connection between the axion scale and SUSY breaking scale:
\bea
f_a(t_0) \,\sim\, v(t_0) \,\sim\,  (m_{\rm SUSY}M_{Pl}^n)^{1/(n+1)},
\eea is that
the axion decay constant $f_a(t_I)$ during inflation can be very different from
the present axion decay constant $f_a(t_0)$.
In regard to this, we have again two distinctive possibilities, which
will be discussed in more detail in section \ref{sec:fa}:
\bea
\hspace{-0.5cm}
&a)& \mbox{PQ symmetry restored during inflation with } v(t_I)=0,
\nonumber \\
\hspace{-0.5cm}
&b)& \mbox{PQ symmetry broken at a higher scale with }  v(t_I) \sim
\left(4\pi H_I M_{Pl}^n\right)^{1/(n+1)}.
\nonumber
\eea
In section \ref{sec:3}, we will discuss the cosmological constraints
on the string theoretic QCD axion charged under an anomalous U$(1)_A$ gauge symmetry,
while taking into account this variation of the axion decay constant from the inflationary
epoch to the present universe.
In section \ref{sec:fa}, we examine the symmetry breaking pattern during inflation
in the context of simple SUGRA model involving both a chaotic inflaton sector
and the U$(1)_A$ sector for the QCD axion.

\section{Cosmological constraints}\label{sec:3}

The QCD axion is subject to various cosmological constraints depending on whether
the PQ symmetry is restored or not in the early universe.
Let us start with the case where the PQ symmetry is restored during inflation:
\bea
v(t_I)=0.
\eea
In this case, the domain-wall number $N_{\rm DW}$ should be equal to one since otherwise
domain walls formed during the QCD phase transition will overclose the universe.
Even for $N_{\rm DW}=1$, axionic strings are formed during the PQ phase transition, and
develop into a network of strings attached by domain walls during the QCD phase transition.
Then dark matter axions are produced from the annihilations of these topological defects,
as well as from the coherent oscillation of misaligned axion field.
Putting these together, one finds that the relic axion mass density at present is given by
\bea
\label{axion-PQ-restoration}
{\Omega_a} h^2 &=&
\left(\Omega_{\rm mis}+\Omega_{\rm string} +\Omega_{\rm wall}\right)h^2
\nonumber \\
&\simeq & \Big( 0.58+(2.0\pm 1.0) + (5.8\pm 2.8)\Big)
\times 
\left(\frac{\Lambda_{\rm QCD}}{400\, {\rm MeV}}\right) \left(\frac{f_a(t_0)}{10^{12}\,
{\rm GeV}}\right)^{1.19},
\eea
where we have used  the results of the recent numerical simulation for the axion production
from the collapsing string and wall system~\cite{Hiramatsu:2012gg},\footnote{
Axion radiation by the string-wall system is determined mostly by the string and wall tensions
given by $\mu_{s}\sim f_a^2\ln \left(m_\phi t\right)$ and $\sigma_{w}\sim m_a f_a^2$,
where $m_\phi$ is the mass of the PQ breaking field.
It was assumed that $m_\phi \sim f_a$ in Ref.~\cite{Hiramatsu:2012gg}, while in our case
$m_\phi\sim m_{\rm SUSY}$.
This may cause a non-negligible change of the axion mass density produced by
the string-wall system.
As it does not change the order of magnitude of the axion mass density, we ignore
this point in the present discussion.
}
together with the root-mean-square value of the axion misalignment angle
$\langle \theta_0^2\rangle \simeq 1.85\times \pi^2/3$, which takes into account
the anharmonic factor $1.85$.
Combined with astrophysical constraints, the condition $\Omega_a\leq \Omega_{\rm DM}$
determines the allowed range of the axion decay constant as
\bea
10^9 \,{\rm GeV} \,<\, f_a(t_0) \,<\, (2-4)\times 10^{10}\,{\rm GeV},
\eea
when the PQ symmetry was restored during inflation, where
$\Omega_{\rm DM}\simeq 0.25$ denotes the total dark matter energy density.
Applying this to the previously discussed scheme generating the axion scale as
\bea
f_a(t_0)\sim (m_{\rm SUSY}M^n_{Pl})^{1/(n+1)},\eea
we are led to either  the axion scale SUSY ($n=0$) or the TeV scale SUSY ($n=1$),
\bea
n=0: && m_{\rm SUSY}\,\sim\, 10^9-10^{10}\,{\rm GeV},
\nonumber \\
n=1: && m_{\rm SUSY} \,\sim\, 10^3\,{\rm GeV}.
\nonumber
\eea

Another, presumably more interesting, scenario is that the PQ symmetry is broken during
inflation at a scale much higher than the present axion scale.
In such case, there are no topological defects, but the axion can still cause cosmological
problems since during inflation it acquires quantum fluctuations
\bea
\delta a(t_I) = \frac{H_I}{2\pi},
\eea
for the canonically normalized axion field during inflation, $a(t_I)=f_a(t_I)\theta_a$.
In models with anomalous U$(1)_A$ gauge symmetry, one combination of the GS axion
$\theta_{\rm st}$ and the matter field axion $\theta_\phi={\rm arg}(\phi)$ becomes
the longitudinal component of the massive U$(1)_A$ gauge boson having a mass much
larger than $H_I\simeq 10^{14}$~GeV, while the other U$(1)_A$-invariant combination can be
identified as the QCD axion.
The fraction of each component in the QCD axion changes with time, and the main component
during and after inflation are different if $v(t_0)\ll f_{\rm st} \lesssim v(t_I)$.
The axion fluctuation around the average misalignment at the moment of coherent
oscillation is given by
\bea
\label{axion-fluctuation}
\langle \delta \theta^2 \rangle
= \left\langle \frac{\delta a^2(t_0)}{f^2_a(t_0)} \right\rangle
= \left(\frac{H_I}{2\pi f_a(t_I)}\right)^2,
\eea
where $\delta\theta=\delta a(t_0)/f_a(t_0)=\delta a(t_I)/f_a(t_I)$ has been used.
For axion models with U$(1)_A$, the ratio between the axion
scales during and after inflation is estimated to be
\bea
\frac{f_a(t_I)}{f_a(t_0)} \simeq
\frac{v(t_I)}{v(t_0)}
\left(\frac{
(8\pi^2\delta_{\rm GS} f_{\rm st})^2 + v^2(t_0)}
{(8\pi^2\delta_{\rm GS} f_{\rm st})^2+v^2(t_I)}\right)^{1/2},
\eea
where we have used the relations (\ref{canonical-axion}) and (\ref{fa-axion}).
Note that the expectation value of
$f_{\rm st}=\sqrt{2\partial^2_\tau K_0}M_{Pl}/8\pi^2$ does not change significantly during and
after inflation as the GS modulus $\tau$ is stabilized by the U$(1)_A$
$D$-term potential at a value near the point of vanishing
FI term in both periods, with a superheavy mass $M_\tau \sim \delta_{\rm GS}M_{Pl}$.
It is also important to note that $f_a(t_I)/f_a(t_0)$ is always smaller than about
$f_{\rm st}/v(t_0)$ for $v(t_0)\ll v(t_I)$.

The axion field is uniform, $a(t_0)=f_a(t_0)\theta_0$, at the classical level
throughout the whole observable universe if the PQ symmetry were broken during inflation.
In addition to this misalignment, there are axion fluctuations $\delta a(t_0)$ induced
during inflation, which are subject to various cosmological constraints.
Let us summarize the constraints, which depend on the values of
$\theta_0$, $f_a(t_0)$, $f_a(t_I)$, $H_I$, and $\Omega_a/\Omega_{\rm DM}$.
We first have the obvious condition:
\bea
\label{axion-abundance}
\frac{\Omega_a}{\Omega_{\rm DM}} \simeq
0.11 ( \theta^2_0 + \langle \delta \theta^2 \rangle
)
\left(\frac{\Lambda_{\rm QCD}}{400\, {\rm MeV}}\right)
\left(\frac{f_a(t_0)}{10^{11}{\rm GeV}}\right)^{1.19}
\,\leq\, 1,
\eea
neglecting anharmonic effects, which become important if the axion initial
position is very close to the hilltop of the potential.
Note that one cannot avoid the contribution from the axion fluctuation
$\langle \delta \theta^2 \rangle \propto H_I^2$.
The QCD axion obtains mass after the QCD phase transition.
Then its fluctuations lead to isocurvature density perturbations of axion dark
matter and also to non-Gaussianity \cite{works-on-axion-iso}, which are strongly constrained
by the observed CMB power spectrum.

The power spectrum of axion isocurvature perturbations is given by \cite{axion-cosmology2}
\bea
{\cal P}_S &=&
2 \left(\frac{\Omega_a}{\Omega_{\rm DM}}\right)^2
\frac{2\theta^2_0 + \langle \delta \theta^2 \rangle
}{(\theta^2_0 + \langle \delta \theta^2 \rangle)^2
}
\langle \delta \theta^2 \rangle
\nonumber \\
&\simeq&
\frac{0.44}{x}
\left(\frac{\Omega_a}{\Omega_{\rm DM}}\right)
\left(\frac{H_I}{2\pi f_a(t_I) }\right)^2
\left(\frac{f_a(t_0)}{10^{11}{\rm GeV}}\right)^{1.19},
\eea
where we have used the relation (\ref{axion-abundance}) with $\Lambda_{\rm QCD}\simeq 400$~MeV,
and $x$ is defined by
\bea
x &\equiv&
2\frac{\theta^2_0 + \langle \delta \theta^2 \rangle
}{
2\theta^2_0 + \langle \delta \theta^2 \rangle
}
\,=\, 1\,\mbox{--}\,2.
\eea
The isocurvature power
is constrained by the Planck observations~\cite{Ade:2013uln}
to be
\bea
\frac{{\cal P}_S}{{\cal P}_\zeta} < 0.041 \,\,(95\%\,{\rm C.L.}),
\eea
where ${\cal P}_\zeta \simeq 2.19\times 10^{-9}$ is the power spectrum of the curvature
perturbations.
Then the isocurvature constraint reads
\bea
\label{iso-cond}
\left(\frac{\Omega_a}{\Omega_{\rm DM}}\right)
\left(\frac{f_a(t_0)}{10^{11}{\rm GeV}}\right)^{1.19}
&<&
2 \times 10^{-10}\,
x \left(\frac{H_I}{2\pi f_a(t_I)}\right)^{-2}.
\eea
%
In addition, there appears non-Gaussianity in isocurvature fluctuations~\cite{Kawasaki:2008sn},
and the experimental bound is roughly translated into
\bea
\label{NG-cond}
\left(\frac{\Omega_a}{\Omega_{\rm DM}}\right)^{1/2}
\left(\frac{f_a(t_0)}{10^{11}{\rm GeV}}\right)^{1.19}
&\lesssim& 10^{-6}
\left(\frac{H_I}{2\pi f_a(t_I)}\right)^{-2}.
\eea
Finally, the existence of the average misalignment angle $\theta_0$ contributing
to the relic axion abundance leads to the condition
\bea
\label{theta-cond}
\frac{\Omega_a}{\Omega_{\rm DM}}
&\gtrsim&
0.11
\left(\frac{H_I}{2\pi f_a(t_I)}\right)^2
\left(\frac{f_a(t_0)}{10^{11}{\rm GeV}}\right)^{1.19},
\eea
taking $\Lambda_{\rm QCD}\simeq 400$~MeV.
When combined with this, the isocurvature constraint (\ref{iso-cond}) puts a severe
upper bound on the axion decay constant at present:
\bea
\label{combined-cond}
f_a(t_0) &<&
7.1
\times 10^{13}\,{\rm GeV}
\left(\frac{H_I}{10^{14}{\rm GeV}}\right)^{-1.68}
\left(\frac{f_a(t_I)}{10^{17}{\rm GeV}}\right)^{1.68},
\eea
which applies independently of the value of $\Omega_a$.
Similarly, the constraint (\ref{NG-cond}) from non-Gaussianity can be combined with
(\ref{theta-cond}) to give a upper bound on $f_a(t_0)$, however the resulting bound
is always weaker than the bound (\ref{combined-cond}) from isocurvature perturbations.
This implies that, it is the isocurvature constraint that determines the cosmologically
viable range of the axion mass density and decay constant.

Before going further, let us discuss anharmonic effects, which have been neglected so
far.
The axion abundance produced from the coherent oscillation is enhanced if the initial
axion position is close to the hilltop~\cite{Visinelli:2009zm,Kobayashi:2013nva,anharmonic-effect},
where the axion potential is not approximated by a quadratic potential.
Such effects can be included by taking
\bea
\langle \theta^2 \rangle \,\to\, \langle F(\theta^2) \theta^2 \rangle,
\eea
in the relation for the axion density (\ref{axion-abundance}), with
$F$ given by \cite{Visinelli:2009zm}
\bea
F(z) \simeq \left(\ln\left(\frac{e}{1-z/\pi^2}\right)\right)^{1.19},
\eea
for $0\leq z <\pi^2$.
The anharmonicity factor $F(z)$ increases from unity as $z$ increases.
The axion contribution to isocurvature density perturbations is also enhanced
as the initial position approaches the hilltop.
One can estimate it using that the axion abundance is proportional to
$F(\theta^2)\theta^2$ \cite{Kobayashi:2013nva},
\bea
{\cal P}_S  =
\left(1 +
\frac{F^\prime(\theta^2_0)}{F(\theta^2_0)}
\right)^2
\times
4\left(\frac{\Omega_a}{\Omega_{\rm DM}}\right)^2
\frac{\langle \delta \theta^2 \rangle}{\theta_0^2}
\left( 1
+ {\cal O}\left( \frac{\langle \delta \theta^2 \rangle}{\theta_0^2}\right)
\right),
\eea
where $F^\prime\equiv z dF(z)/dz$.
Thus the isocurvature perturbation is enhanced approximately by the factor, $(1+F^\prime/F)^2$,
for small fluctuations $\langle \delta \theta^2 \rangle \ll \theta^2_0$.
Including this enhancement factor, one finds that the upper bound on $\Omega_a$
from the isocurvature constraint is smaller than the value obtained by the relation
(\ref{iso-cond}).

Obviously high scale inflation puts strong constraints on the possible range of the axion scale
and relic abundance, which may be satisfied by having a larger decay constant during
inflation~\cite{Linde:1991km}.
As will be discussed in more detail in section \ref{sec:fa}, in models with anomalous U$(1)_A$
gauge symmetry, one can easily obtain
\bea
f_a(t_I) \,\sim\, f_{\rm st}=
\frac{1}{8\pi^2}\sqrt{2\frac{\partial^2 K_0}{\partial \tau^2}}M_{Pl}
\,=\, {\cal O}(10^{-1}-10^{-2})\times M_{Pl},
\eea
with
\bea
v(t_I) \sim (\sqrt{8\pi^2}H_IM_{Pl}^n )^{1/(n+1)}
\gtrsim \, f_{\rm st},
\eea
for a reasonable range of model parameters and $H_I\sim 10^{14}$~GeV.
Then the axion fluctuation (\ref{axion-fluctuation}) is suppressed by
a factor $v(t_0)/f_{\rm st}$ insensitively to the precise value of $v(t_I)$,
relaxing the associated cosmological constraints.

\begin{figure}[t]
\begin{center}
\begin{minipage}{16.4cm}
\centerline{
{\hspace*{0cm}\epsfig{figure=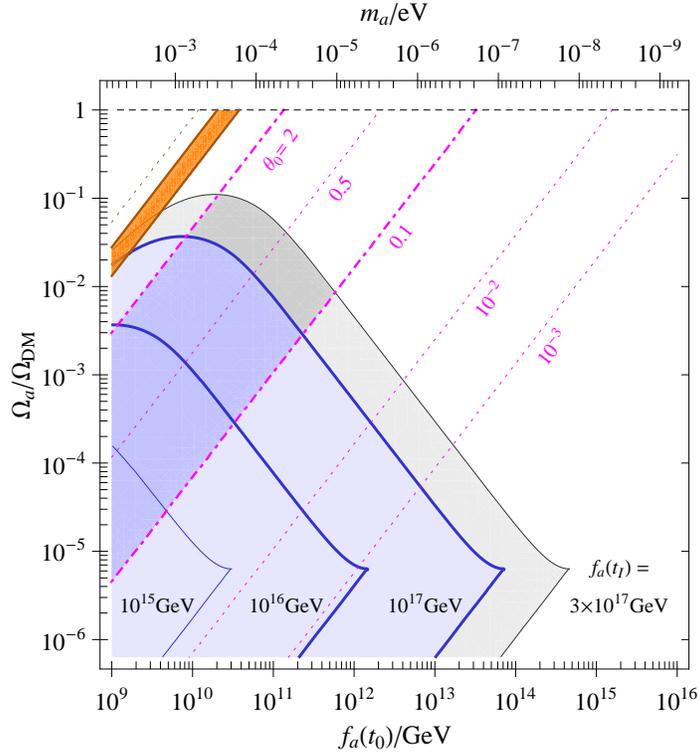,angle=0,width=9.2cm}}
}
\caption{
The axion decay constant (or, equivalently, the axion mass)
and the axion fraction of the dark matter energy density in
the present universe, which are consistent with cosmological constraints.
The orange band is the allowed region when the PQ symmetry is restored
during inflation or reheating, for which the domain-wall number should
be equal to one.
For the case that the PQ symmetry were broken during and after inflation,
there are no topological defects, but the axion abundance and decay constant are severely
constrained by the bound on isocurvature density perturbations.
These cosmological constraints are relaxed in models with anomalous U$(1)_A$ gauge
symmetry since the axion decay constant during inflation can be much higher  than
the present value.
For the inflation scale $H_I=10^{14}$~GeV, the shaded region bounded by the solid lines for
$f_a(t_I)=10^{15},\,10^{16},\,10^{17},\,3\times 10^{17}$~GeV
are cosmologically allowed.
We also show the average axion misalignment angle $\theta_0$
in dot-dashed and dotted lines.
}
\label{fig:axion}
\end{minipage}
\end{center}
\end{figure}

Fig.~\ref{fig:axion} shows the region consistent with the cosmological constraints
in the plane of $f_a(t_0)$ (or, equivalently, the axion mass $m_a$) and
$\Omega_a/\Omega_{\rm DM}$, for $H_I=10^{14}$~GeV.
The orange band, which is obtained from the relation (\ref{axion-PQ-restoration}),
shows the allowed range for the case that the PQ symmetry is restored during inflation
or reheating.
In such case, the axion domain wall number should be equal to one.
On the other hand, if the PQ symmetry remains broken during and after inflation,
the model is free from topological defects, but the axion fluctuations produced during
inflation put strong constraints.
The left side of the solid curves represents the allowed region for
$f_a(t_I)=10^{15},\,10^{16},\,10^{17},\,3\times 10^{17}$~GeV, from
left to right, respectively.
The allowed region becomes larger as $f_a(t_I)$ increases, but the axion cannot account
for the total abundance of dark matter for $f_a(t_I)$ smaller than the Planck scale
\cite{Higaki:2014ooa}.
In models with anomalous U$(1)_A$, we have
$f_a(t_I)\sim f_{\rm st}={\cal O}(10^{16}$--$10^{17})$~GeV
for $v(t_I)\gtrsim f_{\rm st}$, for which the axion fluctuation is suppressed by
$f_a(t_0)/f_a(t_I) \sim (m_{\rm SUSY}M^n_{Pl})^{1/(n+1)}/f_{\rm st}$.
We also show the result for a fixed value of the misalignment angle,
$\theta_0=2,\,0.5,\,0.1,\,10^{-2},\, 10^{-3}$ in the magenta dot-dashed and dotted lines
for $f_a(t_I)=10^{16}$~GeV.
If one takes a larger value of $f_a(t_I)$, the dot-dashed lines move to the right,
but only slightly for $\theta^2_0\gg (H_I/2\pi f_a(t_I))^2$.
Here we have included the full anharmonic effects by taking the initial condition,
$|1-\theta_0/\pi| \lesssim H_I/f_a(t_I)$, so that the axion does not pass over
the hilltop of the potential in the presence of fluctuations.

We close this section by summarizing  the cosmologically viable values of the axion
decay constant and relic abundance.
The value of $f_{\rm st}= \sqrt{2\partial^2_\tau K_0} M_{Pl}/8\pi^2$ lies in the range
between about $10^{16}$ and $10^{17}$~GeV, independently of the details of moduli
stabilization.
Then the isocurvature constraint requires the axion decay constant at present to be
\bea
\label{fa-bound}
f_a(t_0) <
%
7.1
\times 10^{13}\,{\rm GeV}
\left(\frac{f_{\rm st}}{10^{17}{\rm GeV}}\right)^{1.68},
\eea
as can be seen from the relation (\ref{combined-cond}) by taking $f_a(t_I)\simeq f_{\rm st}$,
and also from Fig.~\ref{fig:axion}, where
$f_a(t_0)$ around $10^{14}$~GeV would require a severe fine-tuning of
the axion misalignment angle $\theta_0$.
On the other hand, the natural value of $\theta_0$ would be of the order of unity, which
corresponds to the dark shaded region between the two dot-dashed lines.
For such natural value of $\theta_0$, the isocurvature constraint is translated into
\bea
f_a(t_0) < 6 \times 10^{10}\,{\rm GeV}
\left(\frac{\theta_0}{0.5}\right)^{-0.84}
\left(\frac{f_{\rm st}}{10^{17}{\rm GeV}}\right)^{0.84}.
\eea
Hence, in the case that the PQ symmetry is broken during inflation with a misalignment
angle $\theta_0={\cal O}(1)$, the QCD axion
is expected to have a decay constant in the range around
$10^9-10^{11}$~GeV, while composing up to $0.1-10$~\% of the total dark matter
energy density.
If this were the case, one is again led to either the axion scale SUSY ($n=0$),
\bea
m_{\rm SUSY} \sim  v(t_0) \sim 10^9-10^{11} \,\, {\rm GeV},
\eea
or the low scale SUSY ($n=1$) with
\bea
m_{\rm SUSY}\sim \frac{v^2(t_0)}{M_{Pl}} \sim 10^3-10^4\,\, {\rm GeV},
\eea
because the axion scale is determined by $v(t_0)\sim (m_{\rm SUSY}M^n_{Pl})^{1/(n+1)}$,
and sizable suppression of axion fluctuations is achieved for $v(t_I)\sim f_{\rm st}$
(see also section \ref{sec:fa}).
Although unnatural, $\theta_0$ may have a value much smaller than one,
which would allow a larger axion decay constant as (\ref{fa-bound}).
In this case, the QCD axions constitute only a negligibly small fraction of the observed
dark matter energy density.

\section{Axion decay constant during and after inflation}\label{sec:fa}

In this section, we examine the PQ symmetry breaking both at present and during inflation
in the context of simple supergravity model involving the $U(1)_A$ and inflaton sectors.
We begin with a configuration with vanishing FI term,
\bea
\xi_{\rm FI} \propto
\left.\frac{\partial K_0}{\partial \tau}\right|_{\,\tau=\tau_0} = 0,
\eea
where $T=\tau/2+i\theta_{\rm st}$ is the modulus-axion superfield implementing
the GS anomaly cancellation mechanism.
For simplicity, we consider a minimal U$(1)_A$ sector involving the vector multiplet
$V_A$, the GS multiplet $T$, and two matter fields $\phi_i$ ($i=1,2$) with opposite
sign of U$(1)_A$ charges.
Then, the K\"ahler potential and superpotential of the U$(1)_A$ sector can be expanded around
the configuration $T=\tau_0/2$ and $\phi_i=0$ as
\bea
K &=&
\frac{M^2_{Pl} \partial^2_\tau K_0(\tau_0)}{2}
%
(\tau-\tau_0 -\delta_{\rm GS}V_A)^2 +\phi_1^*e^{-V_A}\phi_1
+ \phi^*_2 e^{(n+2)V_A}\phi_2,
\nonumber \\
W &=& \lambda\frac{\phi^{n+2}_1 \phi_2}{M_{Pl}^n},
\eea
where we have assumed that the matter K\"ahler metric are moduli-independent, and the
U$(1)_A$ charges of $\phi_i$ are chosen as $q_1=1$ and $q_2=-(n+2)$.

The $D$-flat direction of the U$(1)_A$ sector is lifted by SUSY breaking effects,
and eventually determines the PQ breaking scale as
\bea
f_a =
\frac{f_{\rm st} v}{\sqrt{(8\pi^2\delta_{\rm GS} f_{\rm st})^2+v^2}},
\eea
where
\bea
f_{\rm st} &=& \frac{\sqrt{2 \partial^2_\tau K_0} }{8\pi^2} M_{Pl},
\nonumber \\
\quad v^2 &=& 2 \sum_i q_i^2 \langle |\phi_i|^2\rangle = \sum_i q^2_i v^2_i,
\eea
with $v_i\equiv \sqrt2\langle|\phi_i|\rangle$.
It is thus important to know how the $D$-flat direction couples to the SUSY breaking
sector in the model.

As a concrete example, we introduce a Polonyi-like field $Z$ for the SUSY breaking at present, and
an additional field $X$ which develops a large SUSY breaking $F$-term during inflation
described by the inflaton superfield $\Phi$.
For a large field inflation within the supergravity framework, we assume an approximate
shift symmetry, $\Phi\rightarrow \Phi+ic$.
Then the K\"ahler potential and superpotential of the SUSY breaking sector are given by
\bea
K_{\rm SB} &=& |Z|^2 - \frac{|Z|^4}{\Lambda^2} +  \frac{1}{2}(\Phi+\Phi^*)^2 + |X|^2,
\nonumber \\
W_{\rm SB} &=& \omega_0 + M^2Z + \mu X \Phi.
\eea
Following Ref.~\cite{Kawasaki:2000yn}, it is assumed that the inflaton sector fields, $\Phi$ and $X$,
are odd under a $Z_2$ symmetry, and their superpotential coupling preserves $R$-symmetry,
but explicitly breaks the shift symmetry of $\Phi$.
Note that inflation is driven along the ${\rm Im}(\Phi)$ direction by the $F$-term potential
of $X$.
In the present universe, the inflaton sector fields are settled at $X(t_0)=\Phi(t_0)=0$, and
SUSY breaking is due to the $F$-term of the Polonyi-like field:
\bea
F^Z \simeq \sqrt3 m_{3/2}M_{Pl},
\eea
where $m_{3/2}$ is the gravitino mass in the present universe with nearly vanishing
cosmological constant.
On the other hand, during inflation, 
SUSY breaking is dominated by
\bea
F^X \simeq \mu \varphi(t_I) \simeq  \sqrt3 H_I M_{Pl},
\eea
where $\varphi={\rm Im}(\Phi)$ is the inflaton field, which takes a value larger
than the Planck scale to implement the inflation.

The potential for the $D$-flat direction is generated from the coupling between
the U$(1)_A$ sector and the SUSY breaking sector, which generically take the form,
\bea
\label{coupling}
\Delta K =
(k |Z|^2 + \kappa |X|^2)(\tau-\tau_0-\delta_{\rm GS}V_A)
+ \frac{ k_i |Z|^2 + \kappa_i  |X|^2}{M_{Pl}^2}\,\phi^*_i e^{-q_iV_A}\phi_i,
\eea
when expanded around $T=\tau_0/2$ and $\phi_i=0$.
After integrating out the $F$-term SUSY breaking by $F^{Z,X}$, the scalar potential relevant to
the stabilization of the $D$-flat direction is given by
\bea
V &\simeq& \frac{g^2_A}{2} D^2_A
+ V_0(\tau)
+ (n+2)^2\lambda^2 \frac{|\phi_1|^{2(n+1)}|\phi_2|^2}{M^{2n}_{Pl}}
+ \lambda^2 \frac{|\phi_1|^{2(n+2)}}{M^{2n}_{Pl}}
\nonumber \\
&&
+\, m^2_1 |\phi_1|^2 + m^2_2 |\phi_2|^2
- \left(\lambda A_\phi \frac{\phi^{n+2}_1 \phi_2}{M^{n}_{Pl}} + {\rm h.c.} \right),
\eea
with
\bea
D_A &=& |\phi_1|^2-(n+2)|\phi_2|^2
-\delta_{\rm GS}
\frac{\partial K_0}{\partial\tau},
\\
\label{V_0}
V_0 &=& e^{K_0}
\left( \frac{M^4}{1+k (\tau-\tau_0)}
+ \frac{\mu^2\varphi^2}{1+\kappa (\tau-\tau_0)} - 3|W_{\rm SB}|^2 \right),
\eea
where $m^2_i$ parameterize the  soft scalar masses generated by the $F$-term
SUSY breaking.
It is clear that the phase of $\phi^{n+2}_1\phi_2$ is fixed by the $A$-term
alone.
Using this, one can always take a field basis such that $A_\phi$
is real and positive.
From the above scalar potential, we find the stationary conditions to be
\bea
\label{min-condition}
\partial_\tau V &\propto& g^2_A D_A -
\frac{1}{\delta_{\rm GS}
\partial^2_\tau K_0}
\left( \frac{\partial V_0}{\partial\tau} +\cdots \right) = 0,
\nonumber \\
\partial_{\phi_1} V &\propto&
|\phi_1| \left(
G_1(|\phi_i|,\tau) - (n+2)\lambda A_\phi \frac{|\phi_1|^n|\phi_2|}{M^n_{Pl}}
\right)= 0,
\nonumber \\
\partial_{\phi_2} V &\propto&
|\phi_2|\, G_2(|\phi_i|,\tau) - \lambda A_\phi \frac{|\phi_1|^{n+2}}{M^n_{Pl}} = 0,
\eea
for $G_i$ given by
\bea
G_1 &=& g^2_A D_A + m^2_1 + (n+1)(n+2)^2 \lambda^2
\frac{|\phi_1|^{2n}|\phi_2|^2}{M^{2n}_{Pl}}
+ (n+2)\lambda^2 \frac{|\phi_1|^{2n+2}}{M^{2n}_{Pl}} + \cdots,
\nonumber \\
G_2 &=& -(n+2)g^2_A D_A + m^2_2
+ (n+2)^2\lambda^2 \frac{|\phi_1|^{2n+2}}{M^{2n}_{Pl}}
+ \cdots,
\eea
where the ellipsis indicates terms of higher order in $|\phi_i|^2$ and $(\tau-\tau_0)$.
Among the three pseudo-scalar fields,
\bea
\theta_{\rm st}={\rm Im}(T),\quad  \theta_1=\arg(\phi_1), \quad
\theta_2=\arg(\phi_2),
\nonumber
\eea
the combination $\theta_2+(n+2)\theta_1$ is stabilized by the
$A$-term, while the other two remain massless.
One of them is absorbed into the U$(1)_A$ gauge boson, and the other corresponds to
the QCD axion.

Let us now examine the vacuum configuration in the present universe
with $X=\Phi=0$, and the resulting axion decay constant.
First, the condition $\partial_\tau V=0$ reads
\bea
g^2_A D_A - \frac{1}{\delta_{\rm GS} \partial^2_\tau K_0} \left(
\frac{\partial K_0}{\partial\tau} - \frac{k}{1+(\tau-\tau_0)k} \right) V_0 = 0,
\eea
where $k$ is the coupling between the GS modulus-axion multiplet $T=\tau/2+i\theta_{\rm st}$
and the Polony-like field $Z$ in the K\"ahler potential (\ref{coupling}).
The SUSY breaking by $F^Z$ cancels the cosmological constant,
implying that $V_0(\tau)$ does not play an important role in stabilizing the modulus $\tau$.
The U$(1)_A$-charged $\phi_i$  are stabilized away from the origin if they obtain
tachyonic soft masses and/or sizable $A$-term.
For instance, if the $A$-term is small as
\bea
|A_\phi|\ll \sqrt{|m^2_i|} \sim m_{\rm SUSY},
\eea
the scalar potential has a minimum at
\bea
v_1(t_0) &\sim&
\left(\frac{m_{\rm SUSY} M^n_{Pl}}{\lambda}\right)^{1/(n+1)},
\nonumber \\
v_2(t_0) &\sim& \frac{A_\phi v_1(t_0)}{m_{\rm SUSY}}
\,\ll\, v_1(t_0),
\eea
while giving a small FI term:
\bea
\xi_{\rm FI}\simeq v^2_1(t_0)/2,
\nonumber
\eea
implying that $\tau$ is fixed at
\bea
\langle \tau \rangle \simeq \tau_0
+ \frac{v^2_1(t_0)}{2\delta_{\rm GS} \partial^2_\tau K_0(\tau_0) M_{Pl}^2}.
\eea
On the other hand, in the opposite limit with
\bea
\sqrt{|m^2_i|} \ll |A_\phi| \sim m_{\rm SUSY},
\eea
the scalar potential has a minimum at
\bea
v_1(t_0) \simeq \sqrt{n+2}\,v_2(t_0) \sim
\left(\frac{m_{\rm SUSY} M^n_{Pl}}{\lambda}\right)^{1/(n+1)},
\eea
with a tiny FI term, $\xi_{\rm FI}\ll v^2_1(t_0)$.
As a result, in both cases, the QCD axion component and its decay constant are determined as
\bea
\label{fa-today}
\frac{1}{8\pi^2\delta_{\rm GS}}
\frac{a(t_0)}{f_a(t_0)} &=&
\frac{\theta_{\rm st}}{\delta_{\rm GS}}
- \frac{v^2_1(t_0)}{v^2(t_0)}\theta_1
+ (n+2)\frac{v^2_2(t_0)}{v^2(t_0)}\theta_2
\sim
- \frac{v^2_1(t_0)}{v^2(t_0)}\theta_1
+ (n+2)\frac{v^2_2(t_0)}{v^2(t_0)}\theta_2,
\nonumber \\
f_a(t_0) &=&
\frac{f_{\rm st} v(t_0)}{\sqrt{(8\pi^2\delta_{\rm GS} f_{\rm st})^2+v^2(t_0)}}
%
\,\sim\,
\left(\frac{m_{\rm SUSY} M^n_{Pl}}{\lambda}\right)^{1/(n+1)},
\eea
where
\bea
f_{\rm st} \simeq \frac{\sqrt{2\partial^2_\tau K_0} }{8\pi^2} M_{Pl},
\quad
v^2 = v^2_1 + (n+2)^2 v^2_2,
\eea
and the last equality in (\ref{fa-today}) holds for
$v(t_0) \ll 8\pi^2\delta_{\rm GS} f_{\rm st}$, i.e. when
\bea
m_{\rm SUSY} \,\ll \,
10^{16-2n}
\left(\frac{\lambda}{1.0}\right)
\left(\frac{\delta_{\rm GS}}{10^{-2}}\right)^{n+1}{\rm GeV}.
\eea

Let us move on to the scalar potential during inflation with the inflaton field
\bea
\varphi(t_I)={\rm Im}(\Phi(t_I)) > M_{Pl}.
\nonumber
\eea
In this period, the inflaton sector generates a large positive vacuum energy
\bea
V(t_I)\, =\, 3 H_I^2 M_{Pl}^2 \,\simeq \, |F^X|^2 \,=\, \mu^2\varphi^2(t_I).
\eea
Note that $V_0(\tau)$ in (\ref{V_0}) is of the order of $H^2_IM^2_{Pl}$, and thus becomes
important in high scale inflation with $H_I\gg m_{3/2}$.\footnote{
This also implies that the scalar potential of a light modulus can be significantly
modified during inflation, which may cause the moduli runaway problem in high
scale inflation~\cite{Kallosh:2004yh}.
In our case, the GS modulus $\tau$ obtains a heavy mass
$M_\tau\sim \delta_{\rm GS}M_{Pl}\gg H_I$
by the U$(1)_A$ $D$-term potential, and thus is free from the runaway problem.
For other moduli, if exist, we simply assume that they also have a supersymmetric
mass heavy enough to be free from the runaway problem.
}
Such a large $V_0$ enhances the $U(1)_A$ $D$-term,
which can be seen from the minimization condition $\partial_\tau V=0$, yielding
\bea
g^2_A D_A \simeq \frac{3}{\delta_{\rm GS}
\partial^2_\tau K_0
} \left(
\frac{|\phi_1|^2-(n+2)|\phi_2|^2}{\delta_{\rm GS} M_{Pl}^2}
-\kappa \right) H^2_I,
\label{result}
\eea
where $\kappa$ is the coupling between the GS modulus-axion superfield $T$ and the SUSY breaking
superfield $X$ in the K\"ahler potential (\ref{coupling}), and we have used the
relation $\xi_{\rm FI}=\delta_{\rm GS}\partial_\tau K_0M_{Pl}^2$.

In fact, this expectation value of the $D$-term plays a crucial role for the determination
of the vacuum value of $\phi_i$ during inflation.
For the coupling (\ref{coupling}), SUSY breaking soft masses of $\phi_i$ during inflation
are given by
\bea
\tilde m_i^2 &=& m^2_i + q_i g^2_A D_A
\nonumber \\
&\simeq&
(1-\kappa_i) \frac{|F^X|^2}{M_{Pl}^2} + q_i g^2_A D_A =
{\cal O}\left((1-\kappa_i)  H_I^2 \right)+{\cal O}
\left(8\pi^2 \kappa H_I^2 \right),
\eea
where we have used the result (\ref{result}) with $\delta_{\rm GS} ={\cal O}(1/8\pi^2)$.
This suggests that, for the parameter region with
\bea
\kappa \sim (1-\kappa_i),
\eea
which is presumably a natural choice, the soft masses
are dominated by the $D$-term contribution, and then the symmetric solution
$\phi_1=\phi_2=0$ can not be a stable solution as $\phi_i$ have an opposite
sign of U$(1)_A$ charges.
We then have
\bea
\label{v-inflation}
v_1(t_I) &\sim& \left(\frac{H_IM_{Pl}^n}{\lambda|\delta_{\rm GS}|^{1/2}}\right)^{1/(n+1)},
\nonumber \\
v_2(t_I) &\sim& \frac{ A_\phi v_1(t_I)}{|\delta_{\rm GS}|^{1/2}H_I},
\eea
with
\bea
\langle \tau \rangle \simeq  \tau_0  + \frac{v^2_1(t_I)}{2\delta_{\rm GS}
\partial^2_\tau K_0(\tau_0) M_{Pl}^2}.
\nonumber
\eea
We note that $|A_\phi| \ll H_I$ in the chaotic inflation under consideration, because
the SUSY breaking field $X$ is odd under $Z_2$, and carries a non-zero $R$-charge,
which results in $v_1(t_I) \gg v_2(t_I)$.
Then the QCD axion component and its decay constant during inflation are determined as
\bea
\label{fa-inf}
\frac{1}{8\pi^2\delta_{\rm GS} }
\frac{a(t_I)}{f_a(t_I)} &=&
\frac{\theta_{\rm st}}{\delta_{\rm GS}}
- \frac{v^2_1(t_I)}{v^2(t_I)}\theta_1
+ (n+2)\frac{v^2_2(t_I)}{v^2(t_I)}\theta_2
\,\sim\,
\frac{\theta_{\rm st}}{\delta_{\rm GS}} - \theta_1,
\nonumber \\
f_a(t_I) &\simeq &
\frac{f_{\rm st} v_1(t_I)}{\sqrt{(8\pi^2\delta_{\rm GS} f_{\rm st})^2+v^2_1(t_I)}}.
\eea

As noticed from the discussion in the previous section,
a larger axion decay constant during inflation makes it easier to satisfy
the constraints on the axion isocurvature perturbation and non-Gaussianity.
On the other hand, $f_a(t_I)$ in our framework is bounded as
\bea
f_a(t_I) \,\sim\,  {\rm Min}\big(\, f_{\rm st}, \,v_i(t_I)  \,\big),
\eea
implying that we need
\bea
v_1(t_I) \gtrsim f_{\rm st}=\frac{\sqrt{2\partial^2_\tau K_0
} }{8\pi^2} M_{Pl}
= {\cal O}(10^{16}-10^{17}) \,\, {\rm GeV}
\eea
to saturate the bound as
\bea
f_a(t_I)\simeq f_{\rm st}.
\eea
Such a large expectation value of U$(1)_A$-charged matter fields can be obtained
in high scale inflation with
\bea
H_I
\gtrsim
10^{15-2n}\,{\rm GeV}\times
\left(\frac{\lambda}{1.0}\right)
\left(\frac{\delta_{\rm GS}}{10^{-2}}\right)^{n+3/2},
\eea
which follows from the relation (\ref{v-inflation}).
The above is indeed the case for $H_I\simeq 10^{14}$~GeV when
\bea
n=0 \,\,\, \mbox{and} \,\,\, \lambda\lesssim 0.1, \quad \mbox{or} \quad
n\geq 1 \,\,\, \mbox{and} \,\,\, \lambda\lesssim 1.
\nonumber
\eea

Finally we note that, to restore the PQ symmetry, the modulus coupling to
the inflaton sector should be suppressed as
\bea
|\kappa| \lesssim \delta_{\rm GS} \partial^2_\tau K_0 |1-\kappa_i|,
\eea
which means that the GS modulus-axion superfield $T$ is significantly more sequestered
from the SUSY breaking in the inflaton sector  than the U$(1)_A$-charged matter fields.
In addition, we need to arrange the model parameters to make
$\tilde m^2_i = m^2_i + q_ig^2_A D_A$ positive for both $\phi_1$ and $\phi_2$.

\section{Conclusions}\label{sec:5}

In this paper, we have examined the cosmological constraints on string
theoretic QCD axion in the light of the recent PLANCK and BICEP2 results.
We were focusing  on models with anomalous U$(1)_A$ gauge symmetry, which admit
a supersymmetric solution with vanishing Fayet-Illiopoulos (FI) term $\xi_{\rm FI}=0$,
as such models can be realized in many of the known compactified string theories, while
being consistent with all the known cosmological constraints for
a certain range of model parameters.

If the QCD axion is charged under U$(1)_A$, the axion decay constant is determined
essentially by the vacuum expectation values of U$(1)_A$ charged matter fields $\phi$.
To have a phenomenologically viable axion scale, the supersymmetric solution
$\xi_{\rm FI}=\phi=0$ should be destabilized by a tachyonic SUSY breaking mass of $\phi$,
which would result in an intriguing connection between the axion scale and the SUSY breaking
soft masses in the present universe:
$f_a(t_0) \sim (m_{\rm SUSY}M_{Pl}^n)^{1/(n+1)}$ ($n\geq 0$).
We note that such models can have rich symmetry breaking patterns during inflation,
and therefore allow a certain range of the model parameters compatible with strong cosmological
constraints.

If the modulus-axion superfield implementing the Green-Schwarz (GS) anomaly cancellation
mechanism is {\it not} sequestered from the SUSY breaking by the inflaton sector,
the U$(1)_A$-charged matter fields develop a large expectation value
$\langle \phi(t_I)\rangle\sim (\sqrt{8\pi^2} H_I M_{Pl}^n)^{1/(n+1)}$ during inflation,
due to the tachyonic soft scalar mass $m_\phi^2\sim -8\pi^2 H_I^2$ induced by the U$(1)_A$ $D$-term.
This makes it possible that the model is free from the axion domain wall problem,
while satisfying the severe constraints on isocurvature density perturbations for the axion scale
and relic abundance depicted in Fig.~\ref{fig:axion}.
If one allows a fine-tuning of the classical axion misalignment angle $\theta_0$,
then the axion scale in the range $10^9 \, {\rm GeV} < f_a(t_0) < 5\times 10^{13}$~GeV
is cosmologically viable for a reasonable choice of the model parameters. On the other hand,
for $\theta_0={\cal O}(1)$, the allowed range is reduced to
$10^9 \, {\rm GeV} < f_a(t_0) < 10^{11}$~GeV, with the relic axions composing up to
$0.1$--$10$~\% of the total dark matter energy density.

On the other hand, if the dilaton-axion superfield for the GS mechanism is sequestered from
the SUSY breaking by the inflaton sector, it is possible that the PQ symmetry is restored
during inflation with $\langle\phi(t_I)\rangle =0$.
Such scenario is obviously free from the isocurvature constraint, but is subject to
the domain-wall constraint $N_{\rm DW}=1$.
Furthermore, if one adopts the recent numerical simulation for the axion production by
the annihilations of axionic stings and domain walls for the case of $N_{\rm DW}=1$,
one finds that only a narrow range of the axion decay constant,
$10^9 \, {\rm GeV} < f_a(t_0) < (\mbox{a few})\times 10^{10}$~GeV, is allowed.

\end{document}